\documentclass[preprints,article,accept,moreauthors,10pt,a4paper]{mdpi}

\firstpage{1} 
\pubvolume{xx}
\issuenum{1}
\articlenumber{1}
\pubyear{2018}
\copyrightyear{2018}

\usepackage{bm,braket,amsmath,amssymb}

\Title{Photoinduced Dynamics of Commensurate Charge Density Wave
in 1T-TaS$_{2}$ Based on Three-Orbital Hubbard Model}

\Author{Tatsuhiko N. Ikeda $^{1}{}^*$\orcidA{}, Hirokazu Tsunetsugu $^{1}$ and Kenji Yonemitsu $^{2}$}

\AuthorNames{Tatsuhiko N. Ikeda, Hirokazu Tsunetsugu and Kenji Yonemitsu}

\address{%
$^{1}$ \quad Institute for Solid State Physics, University of Tokyo, Kashiwa, Chiba 277-8581, Japan\\
$^{2}$ \quad Department of Physics, Chuo University, Bunkyo, Tokyo 112-8551, Japan}

\corres{Correspondence: tikeda@issp.u-tokyo.ac.jp; Tel.: +81-4-7136-3442}

\abstract{
We study the coupled charge-lattice dynamics in the commensurate
charge density wave (CDW) phase of the layered compound \tas
driven by an ultrashort laser pulse.
For describing its electronic structure, we employ a tight-binding model
of previous studies including the effects of lattice distortion associated with the CDW order.
We further add on-site Coulomb interactions and reproduce an energy
gap at the Fermi level within a mean-field analysis.
On the basis of coupled equations of motion for electrons and the lattice distortion,
we numerically study their dynamics driven by an ultrashort laser pulse.
We find that the CDW order decreases and even disappears during the laser irradiation
while the lattice distortion is almost frozen.
We also find that the lattice motion sets in on a longer time scale
and causes a further decrease in the CDW order even after the laser irradiation.
}


\newcommand{\bfig}{\begin{figure}[H]\centering}
\newcommand{\efig}{\end{figure}}
\newcommand{\beq}{\begin{equation}}
\newcommand{\eeq}{\end{equation}}

\newcommand{\dd}{\mathrm{d}}
\newcommand{\ii}{\mathrm{i}}

\newcommand{\tas}{{1T-TaS${}_2$} }

\newcommand{\oz}{$3z^2-r^2$}
\newcommand{\oxy}{$xy$}
\newcommand{\oxx}{$x^2-y^2$}

\newcommand{\up}{\uparrow}
\newcommand{\down}{\downarrow}

\newcommand{\htb}{\hat{H}_\text{TB}^0}
\newcommand{\htblat}{\hat{H}_\text{TB}}
\newcommand{\htot}{\hat{H}_\text{total}}
\newcommand{\hint}{\hat{H}_\text{int}}
\newcommand{\hel}{\hat{H}_\text{el}}

\newcommand{\iorb}{m}
\newcommand{\isp}{\sigma}
\newcommand{\cre}{\hat{c}^\dag}
\newcommand{\ann}{\hat{c}}
\newcommand{\numop}{\hat{n}}

\newcommand{\transfer}{t}
\newcommand{\orben}{d}
\newcommand{\pos}{\bm{r}}
\newcommand{\posz}{\bm{r}^{\mathrm{eq}}}

\newcommand{\vlat}{V_\text{lat}}

\newcommand{\hlat}{H_\text{lat}}
\newcommand{\elc}{K}
\newcommand{\fel}{F_\text{el}}

\newcommand{\elstate}{\Psi}
\newcommand{\dir}{\bm{\epsilon}}
\newcommand{\opump}{\omega_\text{pump}}
\newcommand{\tpump}{T_\text{pump}}
\newcommand{\tini}{t_\text{init}}
\newcommand{\dabs}{\Delta E_\text{abs}}
\newcommand{\dcdw}{\Delta E_\text{CDW}}
\newcommand{\euni}{E_\text{uniform}}
\newcommand{\eini}{E_\text{init}}

\newcommand{\dds}{dd$\sigma$}
\newcommand{\ddp}{dd$\pi$}
\newcommand{\ddd}{dd$\delta$}

\newcommand{\appr}{\sim\!}

\newcommand{\eV}{\,\text{eV}}
\newcommand{\meV}{\,\text{meV}}
\newcommand{\K}{\,\text{K}}
\newcommand{\fs}{\,\text{fs}}
\newcommand{\mvcm}{\,\text{MV/cm}}
\newcommand{\thz}{\,\text{THz}}

\newcommand{\ps}{\,\text{ps}}

\newcommand{\ike}{T.N.I.}
\newcommand{\tsu}{H.T.}
\newcommand{\yo}{K.Y.}

\begin{document}


\section{Introduction}
Transition-metal dichalcogenides (TMDs) are layered compounds that 
are a testbed to study electron correlations and electron-lattice interactions in two dimensions~\cite{Wilson1975,Rossnagel2011}.
In particular, 
\tas shows a variety of phases including charge density waves (CDWs) accompanied by periodic lattice distortions (PLDs)
at low temperatures~\cite{Wilson1974,Scruby1975,Fung1980,Tanda1984,Sakabe2017}.
In the lowest-temperature phase below 180\K,
electrons experience a lock-in to the commensurate CDW (CCDW) that
forms a $\sqrt{13}\times\sqrt{13}$ hexagram structure (see Figure~\ref{fig:DS}),
which has been also referred to as the Star of David.
The CCDW phase is insulating 
as revealed by early resistivity and
susceptibility measurements~\cite{Wilson1975},
and this is confirmed by recent experiments of
angle-resolved photo-emission spectroscopy (ARPES)~\cite{Ang2012}.
This is believed to be a Mott insulator~\cite{Fazekas1979} (see also \cite{Perfetti2005,Ritschel2015} for alternative scenarios),
and the possibility of spin liquid is under debate~\cite{Law2017,Klanjsek2017}.

\tas has also attracted renewed interest for its nonequilibrium phenomena
driven by photoexcitation, and 
the electron-lattice cooperative dynamics has been extensively
studied~\cite{Perfetti2008,Eichberger2010,Ishizaka2011,Hellmann2012}.
Photoinduced phase transitions have been observed 
between the different CDWs in the phase diagram~\cite{Haupt2016}.
Besides, photoexcitation~\cite{Stojchevska2014,Gerasimenko2017,Ravnik2018,Stojchevska2018} and a gate-voltage
pulse~\cite{Yoshida2014,Yoshida2015,Vaskivskyi2016,Yoshida2017} have realized transitions
to metastable states, which are often referred to as hidden states~\cite{Ichikawa2011,Beaud2014,Li2018}.

Theoretical studies for these intriguing nonequilibrium phenomena
are quite limited except phenomenological Ginzburg-Landau theories~\cite{McMillan1976,Moncton1977,Nakanishi1977}.
This is partly because modeling the electronic structure
has not been settled yet
for the CCDW phase in \tas.
Constructing an empirical tight-binding model
dates back to Mattheiss~\cite{Mattheiss1973},
who proposed a three-orbital Hamiltonian for
describing the normal state without PLD.
Smith et. al.~\cite{Smith1985} studied the case in the presence of the PLD
and discussed the band structure well below the Fermi energy.
Rossnagel and Smith~\cite{Rossnagel2006} included the spin-orbit interaction
and showed that this isolates one narrow band near the Fermi energy.
They conjectured the Mott insulating phase of this band,
but it remains open to examine its validity.

In this paper,
we extend the empirical tight-binding model of Rossnagel and Smith~\cite{Rossnagel2006}
by adding the on-site Coulomb interaction,
and obtain a band gap at the Fermi level within a mean-field approach.
The gap formation is in line with their conjecture~\cite{Rossnagel2006}
and consistent with the electronic property of the CCDW phase of \tas.
We then treat the PLD of the hexagram shape as a dynamical variable
and formulate the coupled equations of motion for the electrons and the PLD.
On the basis of these equations,
we study the electron-lattice cooperative dynamics induced by an ultrashort laser pulse
within a time-dependent mean-field approximation.
On a short time scale,
the lattice distortion is almost frozen,
but the laser input induces coherent electron dynamics
and the charge density wave decreases and even disappears.
On a longer time scale,
the lattice motion sets in
and causes further decrease in the charge-density-wave order
even after the laser irradiation.

\section{Model and Band Structures}
 
\subsection{Three-Orbital Tight-Binding Model: Previous Studies} 

We first introduce a model for the phase without PLD, and this
is the tight-binding Hamiltonian used in \cite{Rossnagel2006}.  
This model involves three $d$-orbitals (\oz, \oxx, and \oxy) of Ta ions
on the two-dimensional triangular lattice,
and is defined as 
\beq\label{eq:tb0}
\htb = - \sum_{\langle i,j\rangle}\sum_{\iorb,\iorb',\isp}  \transfer_{\iorb\iorb'}^{0,ij}\cre_{i\iorb\isp}\ann_{j\iorb'\isp} 
+\sum_i (\sum_{\iorb} \orben_\iorb \numop_{i \iorb}+ \xi\hat{\bm{L}}_i\cdot \hat{\bm{S}}_i),
\eeq
where the summation $\sum_{\langle i,j\rangle}$ runs over the nearest neighbor
site pairs.  
Here $\ann_{i\iorb\isp}$ denotes the electron annihilation operator 
at site $i$, in orbital $\iorb$ $(=$ \oz,\ \oxy,\ \oxy),
with spin $\isp$ $(=\up,\down)$, and 
$\numop_{i\iorb}\equiv \sum_{\isp}\cre_{i\iorb\isp}\ann_{i\iorb\isp}$ is
the orbital density operator.  
The operators $\hat{\bm{L}}_i$ and $\hat{\bm{S}}_i$ represent
the orbital and the spin angular momenta, respectively, and 
their expressions are given in Section~\ref{sec:soi}.
The parameters $\orben_\iorb$ are the crystal field energies,
while the hopping integrals $\transfer_{\iorb\iorb'}^{0,ij}$ are 
obtained as linear combinations of the Slater-Koster parameters~\cite{Slater1954,Miasek1957} 
shown in Table~\ref{tab:parameters}.
The spin-orbit interaction energy is set as $\xi=0.3129\eV$.
The number of electrons is one per site,
and the Fermi energy or the chemical potential is determined accordingly.

\begin{table}[H]
\caption{List of  crystal field energies and  Slater-Koster parameters
in our model. All values are shown in the unit of eV. 
}
\centering
\begin{tabular}{cc}
\toprule
\textbf{crystal field}	& \textbf{Slater-Koster Parameters}	\\
\midrule
$\orben_{3z^2-r^2}=1.489$	& \dds\,$=-0.7955$\\
$\orben_{x^2-y^2}=\orben_{xy}=1.339$		& \ddp\,$= 0.2595$\\
 &\ddd\,$=0.1080 $\\
\bottomrule
\end{tabular}
\label{tab:parameters}
\end{table}

The effect of the PLD is taken into account
by changes in the transfer integrals~\cite{Smith1985}.
Let $\posz_i$ denote the position of 
the $i$-th Ta atom in the absence of a PLD, 
while $\pos_i$ is the position when a PLD is present. 
Upon the change in atomic positions, we set 
the transfer integral between sites $i$ and $j$
as 
\beq
\transfer^{ij}_{\iorb\iorb'}(\{\pos_i\}) = \left(\frac{|\posz_i-\posz_j|}{| \pos_i-\pos_j|} \right)^5\transfer^{0,ij}_{\iorb\iorb'},
\eeq 
following the empirical law for the $d$-electrons.
Thus, the Hamiltonian with taking account of a PLD is given by
\beq\label{eq:tb}
\htblat = - \sum_{\langle i,j\rangle}\sum_{\iorb,\iorb',\isp}  \transfer_{\iorb\iorb'}^{ij}(\{\pos_i\}) \cre_{i\iorb\isp}\ann_{j\iorb'\isp} 
+\sum_i (\sum_{\iorb} \orben_\iorb \numop_{i \iorb}+ \xi\hat{\bm{L}}_i\cdot \hat{\bm{S}}_i).
\eeq
Note that we neglect inhomogeneous changes in local potential $\Delta \orben_{i\iorb}(\{\pos_j\})$
associated with the PLD.


Following the previous studies~\cite{Smith1985,Rossnagel2006},
we consider a specific PLD mode of the hexagram shape shown in Figure~\ref{fig:DS}(a).
Here, the unit cell has 6-fold rotation symmetry and
contains 13 sites categorized into three types as shown
in Figure~\ref{fig:DS}(b); 
one ``A'' (red), six ``B'' (blue), and six ``C'' (orange) sites.  
The distances from A to B and C sites are denoted by $r_B$ and $r_C$
and these are parameterized by one parameter $x$ quantifying the amplitude of the PLD as
\beq
r_B = 1 + 0.064 x; \quad r_C=\sqrt{3} + 0.072 x.
\eeq
Here the length unit is the lattice constant 3.36\,\AA\ in the absence of a PLD.
In \cite{Rossnagel2006}, the value $x=-1$ corresponding
to the values $(r_B,r_C)=(0.936,1.66)$ is used to be consistent with experimental results.
We note that $x<0$ $(x>0)$ corresponds to a PLD with shrinking (expanding) hexagrams.
In calculation, we generate all positions $\{\pos_i\}$ for a given $x$ without shifting any A site,
and obtain the transfer integrals $\transfer^{ij}_{\iorb\iorb'}(\{\pos_i\})$
according to the modified distances $|\pos_i-\pos_j|$ for all pairs of nearest neighbors.

\bfig
\includegraphics[width=12cm]{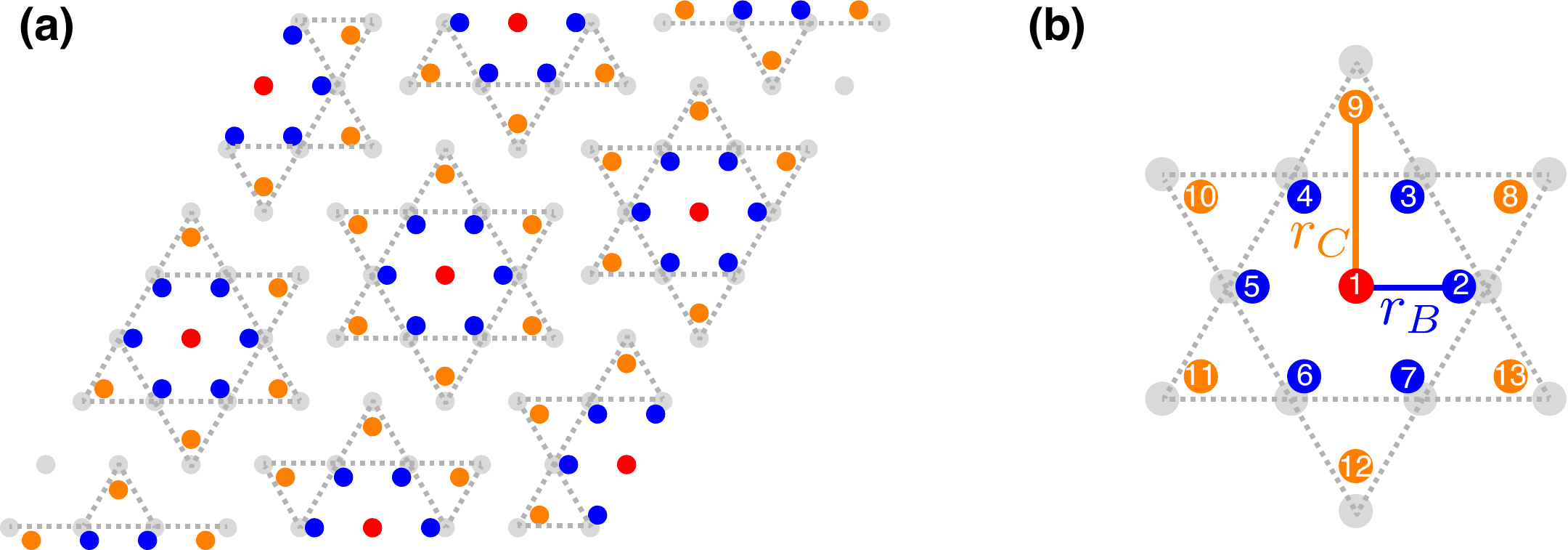}
\caption{
(a) PLD in hexagram configuration in real space (exaggerated). The red, blue, and orange dots show
``A'', ``B'', and ``C'' Ta atoms, respectively.
For reference,  Ta atoms without PLD
are shown by the gray dots.
(b) Parameters $r_B$ and $r_c$ characterizing distorted hexagram configuration.
The configuration is assumed to be symmetric under the $\pi/3$ rotation
and the mirror transformation about the $x$ axis.
The site indices will be used in Section~\ref{sec:pid}.
}
\label{fig:DS}
\efig

The band structure in our tight-binding model is summarized
in Figure~\ref{fig:bands}.
Panel (a) shows the result calculated from 
$\htb$~\eqref{eq:tb0} in the absence of a PLD.
Panel (b) shows the same result but the bands are folded by taking a hexagram as a unit cell.
Panel (c) shows the result calculated from $\htblat$~\eqref{eq:tb} with distortion $x=-1.5$.
This case shows a large band restructuring, and 
one narrow band separates from the others and extends near the Fermi energy.
This isolation becomes clearer for larger $|x|$ with $x<0$.
Note that
we consider the case of $x < -1$, while the value $x=-1$ was used in~\cite{Rossnagel2006}.
This is because the value $|x|=1$ is not large enough
to obtain a splitting of the narrow band
within our treatment of the electron-electron interactions discussed below.

\bfig
\includegraphics[width=15cm]{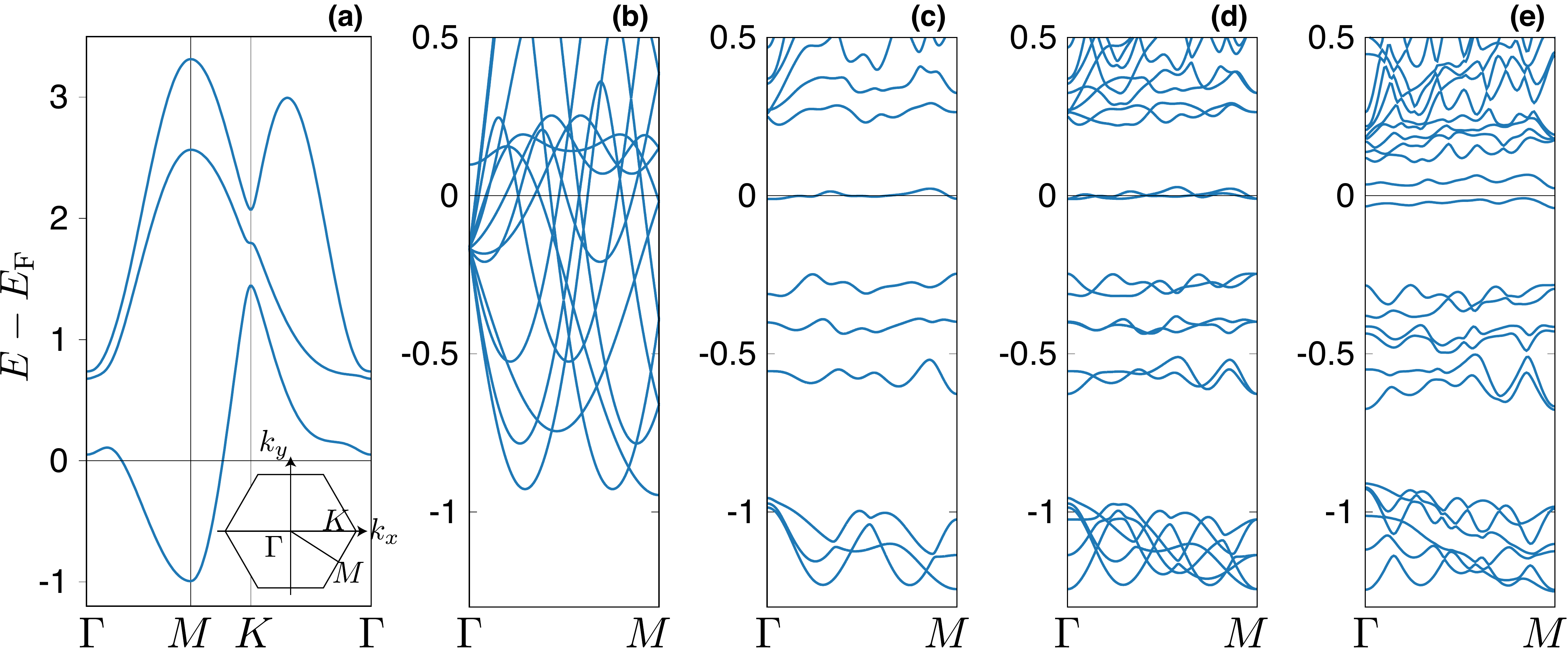}
\caption{
(a) Band structure without PLD or Coulomb interaction. The inset schematically shows the Brillouin zone.
(b) Same result with folded bands obtained by taking each hexagram as a unit cell.
(c) Band structure with PLD ($x=-1.5$) and no Coulomb interaction.
(d) Same result shown with folded bands for two-hexagram unit cell.
(e) Band structure with both  PLD ($x=-1.5$) and  on-site Coulomb interaction ($U=3J=0.816\eV$)
calculated within mean-field approximation~\eqref{eq:mfa} at T=0.
}
\label{fig:bands}
\efig


\subsection{Electron-Electron Interactions}
We now add to $\htblat$~\eqref{eq:tb} the electron-electron interactions of on-site Coulomb type
\beq
\hint = \sum_i \hint^{(i)};\quad
\hint^{(i)} = U\sum_m \hat{n}_{i\iorb \up}\hat{n}_{i\iorb \down}
+U' \sum_{\iorb\neq\iorb'}\hat{n}_{i\iorb\up}\hat{n}_{i\iorb'\down}
+(U'-J)\sum_{\iorb<\iorb',\isp}
\hat{n}_{i\iorb\isp}\hat{n}_{i\iorb'\isp}.\label{eq:hint}
\eeq
Here, we use the standard choice $U'=U-2J$
and ignore the pair hopping terms~\cite{Kanamori1963}.
For simplicity, throughout this paper, we focus on
the typical case of $U=3J$, where the last term in Equation~\eqref{eq:hint} vanishes.
We use a mean-field approach and approximate $\hint^{(i)}$ by
\beq
\hint^{(i)} \simeq U\sum_m\left( {n}_{i\iorb \down}\hat{n}_{i\iorb \up}+n_{i\iorb \up}\hat{n}_{i\iorb \down}  -n_{i\iorb \up}n_{i\iorb \down}  \right)
+U' \sum_{\iorb\neq\iorb'}\left(n_{i\iorb'\down}\hat{n}_{i\iorb\up}+n_{i\iorb\up}\hat{n}_{i\iorb'\down}-n_{i\iorb\up}n_{i\iorb'\down}\right),\label{eq:mfa}
\eeq
where $n_{i\iorb\isp}\equiv \langle \hat{n}_{i\iorb\isp}\rangle$ is determined self-consistently
and we have neglected the orbital off-diagonal contributions $\langle \cre_{i\iorb\isp}\ann_{i\iorb'\isp}\rangle$ for simplicity.

In the mean-field approximation, 
we use the unit cell made of two hexagrams, which accommodates 26 electrons,
and it is important that the electron number is even.  
To simulate the charge distribution in the Mott insulator phase,
we introduce a fictitious magnetic order in the mean-field approximation.  
Yet, this redundant magnetic order is not very harmful for discussing the charge dynamics driven by strong laser fields.
Another drawback is that the translation symmetry is further lowered
than a periodic array of hexagrams, but 
we shall show later that this effect is not very strong.

We compare in Figures~\ref{fig:bands}(d) and (e) the band structure
without and with the electron-electron interaction.  
Panel (d) is plotted for comparison and shows the same energy bands as in (c)
($U=3J=0$), but the bands are now folded corresponding to a two-hexagram unit cell.
Figure~\ref{fig:bands}(e) is the result for $U=3J=0.816\eV$
calculated at zero temperature $T=0$.
The interactions have relatively weak effects at energies well below
the Fermi energy $E_F$.
However, near $E_F$, they cause a splitting of the two bands 
and make the system insulating.  
The band gap at the $\Gamma$ point is 69.2\meV,
which is a few times smaller than the experimental value~\cite{Perfetti2008}.
While the gap size can be reproduced by tuning $U$, $U'$, and $J$, 
we do not go into further detail in this paper.

Let us now examine charge and spin densities.
The results are shown in Figure~\ref{fig:density} for the same parameters $U=3J=0.816\eV$ at $T=0$.
All the hexagrams have the identical charge density distribution: 
the density increases toward the center of each hexagram,
and the 2-fold rotational symmetry around an A site is preserved.  
The 6-fold rotational symmetry is broken due to the doubled 
unit cell, but its asymmetry is so small (about  1\%) that 
this effect is ignorable.
In the doubled unit cell, the two hexagrams have the spin density distribution
with the opposite signs, and this is due to a fictitious Neel order. 
The spin density distribution in each hexagram reflects the profile of
the wave functions of the narrow band in Figure~\ref{fig:bands}(c).
In fact, as was shown in \cite{Rossnagel2006},
these wave functions have a large amplitude near the hexagram center (A site).

\bfig
\includegraphics[width=12cm]{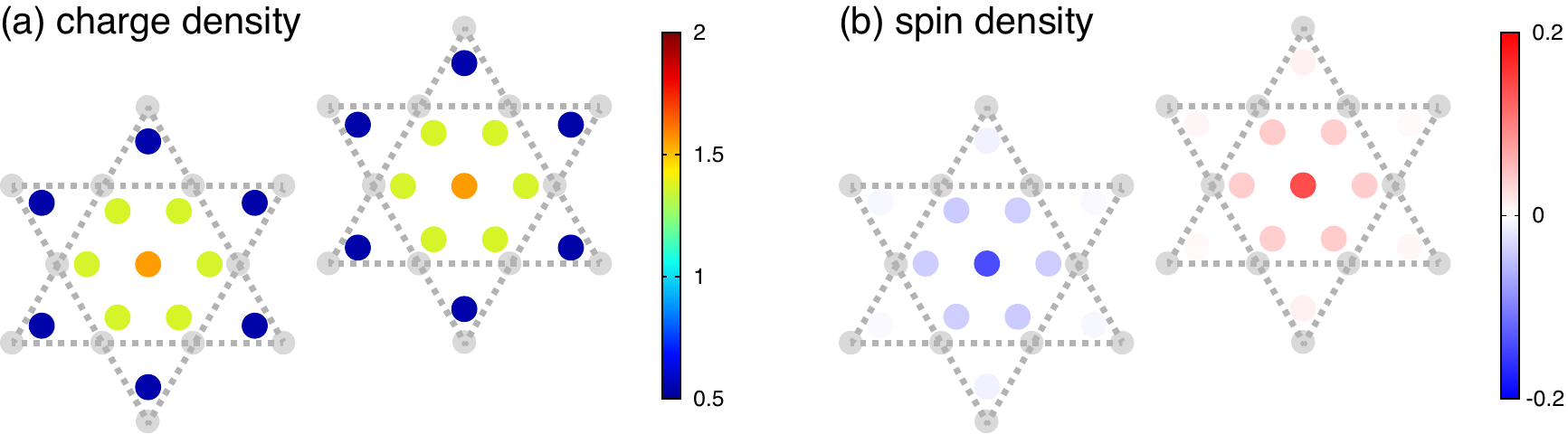}
\caption{
Charge (a) and spin (b) densities in two-hexagram unit cell
in presence of both PLD ($x=-1.5$) and Coulomb interaction ($U=3J=0.816\eV$)
calculated within mean-field approximation at $T=0$.
The charge-density distributions of the two hexagrams are identical, whereas their spin-density distributions have opposite signs.
}
\label{fig:density}
\efig

We emphasize again that the spin polarization appears
due to the technical reason
in treating a Mott insulator by the mean-field approximation~\eqref{eq:mfa}.
In fact, the CCDW phase of \tas is paramagnetic~\cite{Law2017,Klanjsek2017}.
Our aim is not to discuss the spin structure,
but to investigate the charge properties, which play a main role upon strong laser drivings.
For this purpose, we have now set up a reasonable model by adding the Coulomb interaction~\eqref{eq:hint}
to the tight-binding Hamiltonian~\eqref{eq:tb} used in previous studies.

\section{Lattice Degree of Freedom}
To discuss cooperative phenomena,
we now generalize our model and treat the lattice distortion $x$ as a dynamical variable.
We discuss its statics in this section and shall do its dynamics in Section~\ref{sec:pid}.

We parametrize the elastic energy per two-hexagram unit cell accompanied by a PLD as
\beq
\vlat(x) = \frac{\elc_2}{2}x^2 + \frac{\elc_3}{3}x^3 +\frac{\elc_4}{4}x^4,\label{eq:vlat}
\eeq
where $K_2,K_3,$ and $K_4$ are positive parameters so that $\vlat(x)$ is convex.
Equation~\eqref{eq:vlat} is a Taylor series in $x$,
which starts from $x^2$ 
since $x=0$ represents the equilibrium position at high temperatures.
We take account of the $x^3$ term, since the asymmetry between $x<0$ and $x>0$, or shrinking and expanding hexagrams,
become nonnegligible for larger $x$.

We set the parameters $K_2,K_3,$ and $K_4$ so as to satisfy the following three conditions:
(i) the CDW phase appears at $T\sim500\K$,
(ii) $x=-1.5$ in equilibrium at $T=0$,
and (iii) the free energy difference between the CCDW ($x=-1.5$) and the uniform ($x=0$) states
is not very large, or typically less than 1\eV\ at $T=0$ per two-hexagram unit cell.
Here the free energy means the sum of $\vlat(x)$ and 
the electronic free energy $\fel(x)$ calculated for $\htblat(x)+\hint$ within the mean-field approximation~\eqref{eq:mfa}.
The condition (i) determines $K_2$,
and then the conditions (ii) and (iii) do $K_3$ and $K_4$.
We have chosen $K_2=12.1$\eV, $K_3=5.44$\eV, and $K_4=2.54$\eV.

Figure~\ref{fig:fe}(a) shows the electronic energy $\fel(x)$ at $T=0$ and $\vlat(x)$ as functions of $x$.
We note that the free energy coincides with the energy at $T=0$.
The PLD decreases $\fel(x)$ and increases $\vlat(x)$ in both directions $x<0$ and $x>0$.
The competition between these gain and cost results in the minimum of the total free energy $\fel(x)+\vlat(x)$ at $x=-1.5$
as shown in Figure~\ref{fig:fe}(b).
There also exists a very shallow local minimum at $x\sim 0.2$
corresponding to an expanding hexagram.
At higher temperatures above $\sim$800\K,
the total free energy $\fel(x)+\vlat(x)$ has only one minimum at $x=0$ and the CDW is not present.
We note that this transition temperature is probably overestimated
since we have performed the mean-field approximation~\eqref{eq:mfa}
and neglected phonon fluctuations.

\bfig
\includegraphics[width=14cm]{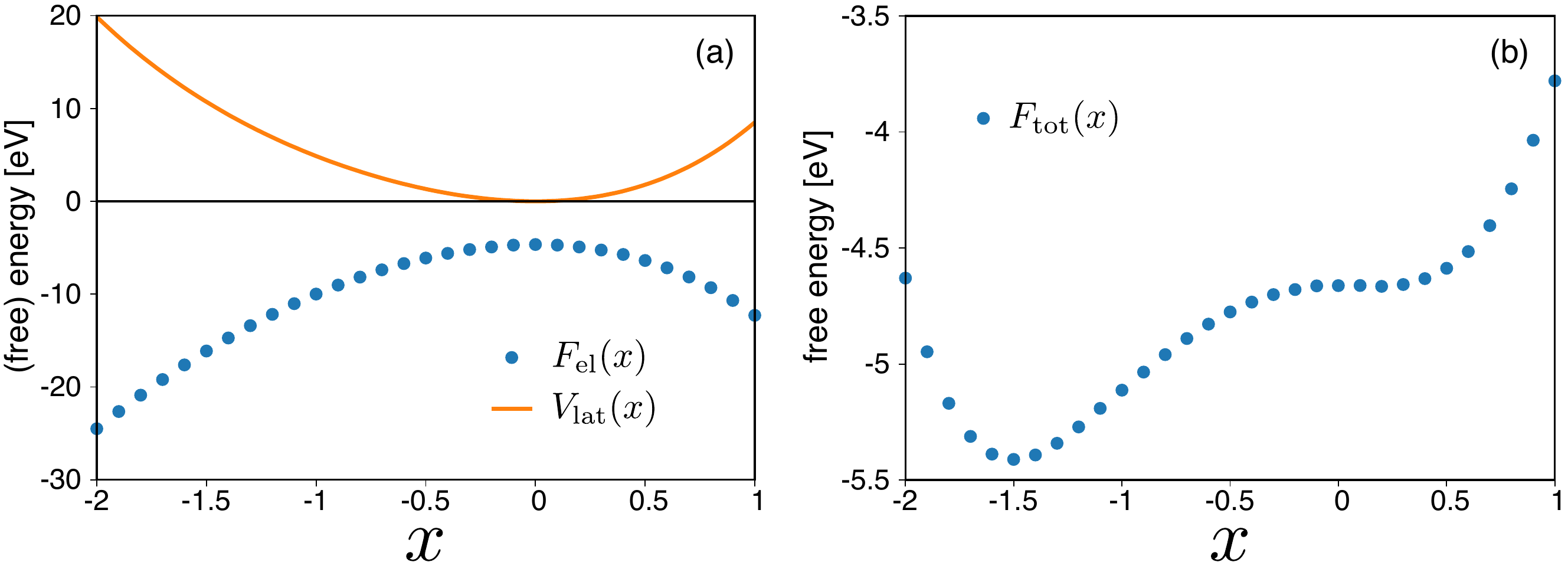}
\caption{
(a) Electronic free energy $\fel(x)$ and elastic energy $\vlat(x)$ accompanied by PLD.
These values are per two-hexagram unit cell and $\fel(x)$ is calculated for $U=3J=0.816\eV$ at $T=0$.
(b) Total free energy $\fel(x)+\vlat(x)$ calculated from data in panel (a).
}
\label{fig:fe}
\efig

We make a remark on the simplification that we have made on the PLD.
We have treated one specific mode of the hexagram shape in lattice distortion,
but the distortion has many more modes.
In fact, \tas shows an incommensurate CDW below 540\K\
and a nearly commensurate CDW below 350\K,
which cannot be described within the present model.
Thus we do not further discuss the free energy at intermediate temperatures,
and shall focus on the CCDW state thus obtained at $T=0$.

\section{Photoinduced Dynamics}\label{sec:pid}
\subsection{Equations of Motion}
To study the photoinduced charge dynamics, we use the coupled equations of motion
for the electrons and the PLD.
We treat the electrons quantum-mechanically
and describe their state by a mean-field wave function $\ket{\elstate(t)}$ at each time $t$.
For the PLD, we classically treat it and describe its coordinate and momentum, $x(t)$ and $p(t)$,
by Hamilton's equations of motion.

The total Hamiltonian is given by
\begin{align}
&\htot = \hel(x(t),t) + \hlat(x(t),p(t)),\label{eq:htotal}\\
&\hel(x(t),t) \equiv \htblat(x(t),t)+\hint,\\
&\hlat(x(t),p(t))\equiv\frac{p(t)^2}{2M}+\vlat(x(t)),
\end{align}
where $M$ is the effective mass of the PLD and its value will be discussed later,
and $\hint$ is treated within the mean-field approximation~\eqref{eq:mfa}.
The explicit time dependence of $\htblat(x(t),t)$ comes from the coupling to the laser electric field,
which oscillates in time.
We assume that the electric field $\bm{E}(t)$ is spatially uniform
and treat it in terms of the vector potential $\bm{A}(t)$ satisfying $\bm{E}(t)=-\dd \bm{A}(t)/\dd t$.
Then $\htblat(x(t),t)$ is obtained from Equation~\eqref{eq:tb}
with the Peierls substitution
\beq
\transfer_{\iorb\iorb'}^{ij}(\{\pos_i(t)\}) \to   \transfer_{\iorb\iorb'}^{ij}(\{\pos_i(t)\}) \exp\left[ \ii (\pos_i(t)-\pos_j(t))\cdot \bm{A}(t) \right],
\eeq
where the elementary charge is set to unity.

The equations of motion are given by
\begin{align}
&\ii \hbar \frac{\dd}{\dd t} \ket{\elstate(t)} = \hel(x(t),t) \ket{\elstate(t)},\label{eq:eom_el}\\
&\frac{\dd x(t)}{\dd t} = \frac{p(t)}{M},\label{eq:lateq1}\\
&\frac{\dd p(t)}{\dd t} =  - \frac{\dd \vlat(x(t))}{\dd x} - \braket{\elstate(t) | \frac{\partial \hel(x(t),t)}{\partial x}  | \elstate(t)}.\label{eq:eom_lat}
\end{align}
In the Schr\"{o}dinger equation~\eqref{eq:eom_el}, we have ignored the c-number term $\hlat(x,p)$ on the right-hand side
since it affects no physical observables.
Equations~\eqref{eq:lateq1} and \eqref{eq:eom_lat} are Hamilton's equations of motion for the PLD.
The right-hand side of Equation~\eqref{eq:eom_lat} represents the force acting on the PLD
and consists of the elasticity and the reaction from the electrons.
We note that, when $\bm{E}(t)=0$, $\hel(x(t),t)$ does not depend explicitly on $t$
and the total energy $\braket{\elstate(t) | \hel(x(t)) | \elstate(t)}+\hlat(x(t),p(t))$ is a conserved quantity.

\subsection{Cooperative Dynamics of Charge and Lattice}
Motivated by recent experiments,
we study the electronic excitation driven by an ultrashort laser pulse.
We take the following form for the vector potential:
\beq
\bm{A}(t) = \dir A_0 \cos(\opump t)  \exp \left( -\frac{t^2}{2\tpump^2}\right).\label{eq:Aform}
\eeq
Here the real parameter $A_0$ denotes the peak amplitude of the vector potential, and
$\dir$ is the polarization vector and assumed to be $\dir=(1,0)$ in this paper.
The results are not sensitive to the direction of $\dir$.
Following the experiment~\cite{Stojchevska2014},
we set the central frequency $\opump$ and the pulse width $\tpump$ so that
$\hbar\opump=1.55\eV$ and $2\tpump=35\fs$.
The profile $A_x(t)$ is shown in Figure~\ref{fig:el_dist}(a).
We note that the peak amplitude of the electric field is approximately given by $E_0=\opump A_0$
since $\tpump \gg 1/\opump=2.4\fs$.
In the following,
we shall use $E_0$ rather than $A_0$ making it easier to compare our results with experimental results.

\bfig
\includegraphics[width=15cm]{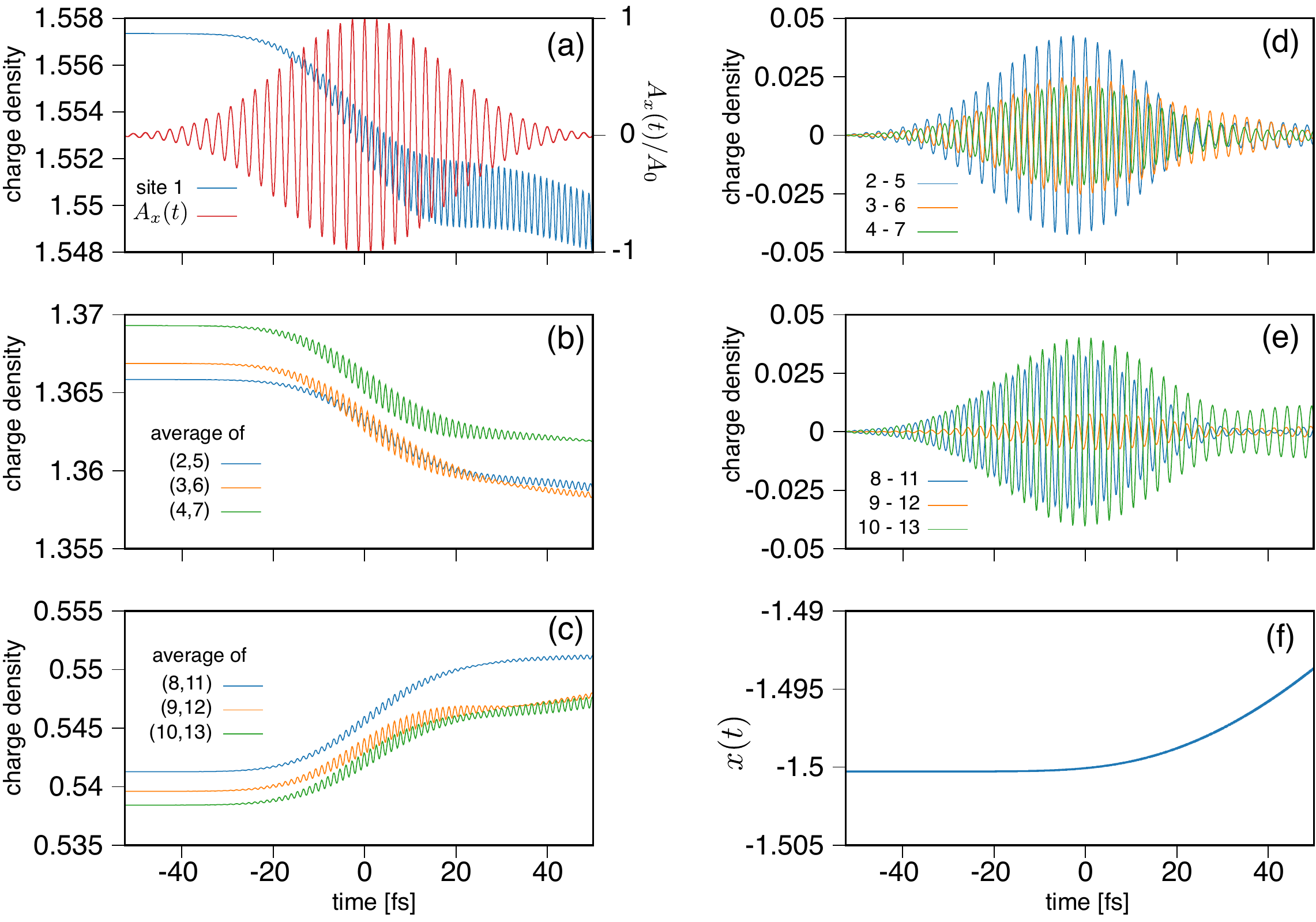}
\caption{
Dynamics results for $E_0=0.92\mvcm$ on short time scale.
Panel (a) shows the charge density at A site (i.e., site 1) (see Figure~\ref{fig:DS}(b) for the site index),
and the input vector potential $A_x(t)$ [Equation~\eqref{eq:Aform}].
Panels (b) and (c) show the time profiles of the average charge density
for the pairs of the ``B'' sites [$(2,5)$, $(3,6)$, and $(4,7)$],
and the ``C'' sites [$(8,11)$, $(9,12)$, and $(10,13)$], respectively.
Panels (d) and (e) show those of charge-density difference in
the pairs of the ``B'' sites [$(2,5)$, $(3,6)$, and $(4,7)$],
and the ``C'' sites [$(8,11)$, $(9,12)$, and $(10,13)$], respectively.
Panel (f) shows the time evolution of the lattice distortion $x(t)$.
}
\label{fig:el_dist}
\efig

The laser electric field induces charge dynamics on short time scale of order $\tpump$.
In Figure~\ref{fig:el_dist}, we show the numerical results of local charge densities at 13 sites
in a hexagram unit for a typical peak amplitude
$E_0=0.92\mvcm$, which is comparable with the experimental condition~\cite{Stojchevska2014}
(see Figure~\ref{fig:DS} for the labeling of sites).
Since we are not interested in the spin density, the charge density averaged between the two hexagrams is plotted.
At our initial time of calculation $t=\tini=-52.5\fs$, 
the charge-density distribution is the one shown in Figure~\ref{fig:density}(a).
As noted before, the 6-fold rotational symmetry is slightly broken down to the 2-fold symmetry.
As time advances, the electric field envelope gradually increases,
and the charge density starts to oscillate in time at each site.

To analyze the charge-density oscillations in detail,
we pair up the six ``B'' sites as $(2,5)$, $(3,6)$, and $(4,7)$,
each of which locates symmetrically about the central A site.
We do the similar pairing for the ``C'' sites as well.
We look into the average and the difference of the charge densities for each site pair,
which are even and odd, respectively, under the 2-fold rotation.
We note that the charge density at site 1 (A) itself is even under this operation.
These even and odd quantities show different kinds of dynamics.
The even quantities shown in panels (a), (b), and (c)
oscillate with frequency $\sim 2\opump$,
whereas the odd ones  (d) and (e) do with frequency $\sim \opump$.
This is because
the vector potential $\bm{A}(t)$ has an odd parity with respect to the 2-fold rotation,
and these two quantities are coupled to $\bm{A}(t)$ in the second and the first orders, respectively.

In addition to those oscillations,
the charge dynamics changes its spatial distribution and
the CDW order decreases on the short time scale of order $\tpump$.
The charge-density imbalance in panels (a-c) decreases during the laser irradiation,
and the local charge density slightly approaches the uniform one.

This CDW order decrease is caused not just by the decrease in the PLD.
Figure~\ref{fig:el_dist}(f) shows that the lattice distortion $x(t)$ remains almost unchanged until $t\sim0$,
while the significant changes have already appeared in the charge density.
Besides, at $0<t<50\fs$, $|x(t)|$ gradually decreases, while the charge dynamics rather slows down.
Therefore, the short-time dynamics is dominated by the electrons driven by the laser field,
and the CDW is rapidly suppressed.

On a longer time scale,
the lattice distortion $x(t)$ plays a main role in dynamics.
We show $x(t)$ up to $t=1000\fs$ in Fig~\ref{fig:lat_long}.
Its time evolution is well approximated by a harmonic oscillation
starting at $t\sim 0\fs$, and its frequency is determined as $2.14\thz$ by a sinusoidal fitting.
This agrees well with the experimentally observed value $2.1\thz$ of the coherent phonon~\cite{Eichberger2010}.
Of course, the frequency depends on the effective mass $M$ of the PLD.
We have set $M=8.00\times10^3\text{u}$ in the unified atomic mass unit
to make the frequency consistent with the experimental result.

\bfig
\includegraphics[width=13cm]{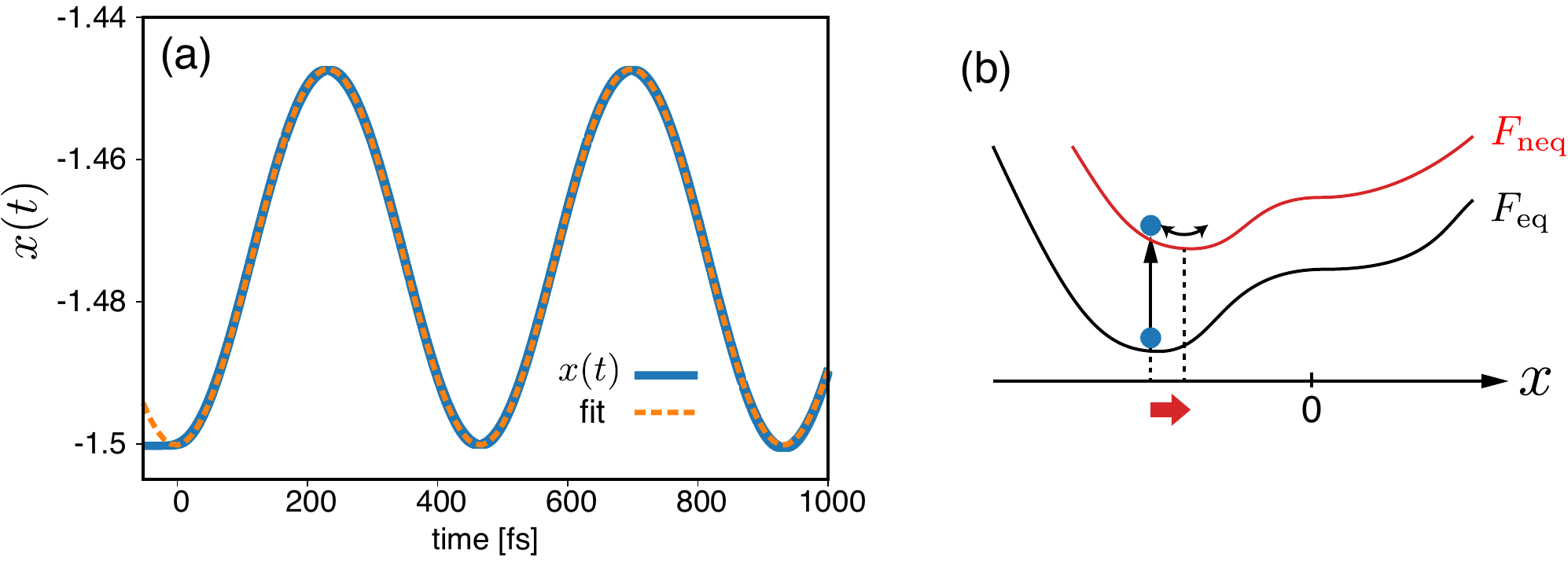}
\caption{
(a) Time profile of lattice distortion $x(t)$ (solid) and its sinusoidal fitting (dashed).
(b) Schematic illustration of mechanism of lattice dynamics $x(t)$.
}
\label{fig:lat_long}
\efig

The lattice dynamics can be interpreted by an instantaneous modulation of the free energy profile
as schematically illustrated in Figure~\ref{fig:lat_long}(b).
In the initial state, the lattice distortion $x$ 
is located at the minimum of the total free energy $F_\text{eq}$.
Then, on the short time scale, the laser drives the electrons to a high-energy state,
and the free energy is changed into a new one $F_\text{neq}$,
whose minimum shifts toward $x=0$.
Since the lattice distortion remains almost unchanged during the driving,
the free energy is not minimum and the distortion $x$
oscillates harmonically around the new minimum of the free energy.

We remark that the harmonic oscillation is not damped in our model since we incorporate no dissipation processes
and the total energy is conserved.
In reality, the time scale of the damping is known to be $\appr 4\ps$.
Thus we do not proceed our analysis after $t\sim1000\fs$, where dissipation would start to play a main role in dynamics.

We briefly summarize the cooperative dynamics revealed by our analysis.
The short-time dynamics is dominated by the electronic degrees of freedom driven by the laser.
The driving rapidly suppresses the CDW order on the time scale of the pulse width, which is $35\fs$ in our analysis.
In this time scale, the PLD is approximately frozen and does not decrease much.
On the long time scale of order $100\fs$, the PLD starts to oscillate at frequency $\appr2\thz$ and will be damped in general through dissipation.
These observations are consistent with the experimental finding~\cite{Eichberger2010},
which observed a delay of the PLD decrease after the CDW order decrease.

\subsection{Melting of Hexagrams}
Now we further increase the laser amplitude
and see how the lattice dynamics changes.
Figure~\ref{fig:stronger} shows
the dynamics of the charge densities averaged over A, B, and C sites
for the larger peak amplitudes $E_0=4.61$ and $23.1\mvcm$
in panels (a) and (b), respectively.
For $E_0=4.61\mvcm$, the CDW order decreases by $\appr20\%$,
which is much enhanced than the case of $E_0=0.92\mvcm$ shown in Figure~\ref{fig:el_dist}.
For $E_0=23.1\mvcm$, the CDW order disappears
and the charge distribution becomes nearly uniform within $\appr20\fs$.
We note that the PLD remains present on this time scale as shown in panel (c).
Only the CDW order of the hexagram shape melts down on the short time scale.

\bfig
\includegraphics[width=15cm]{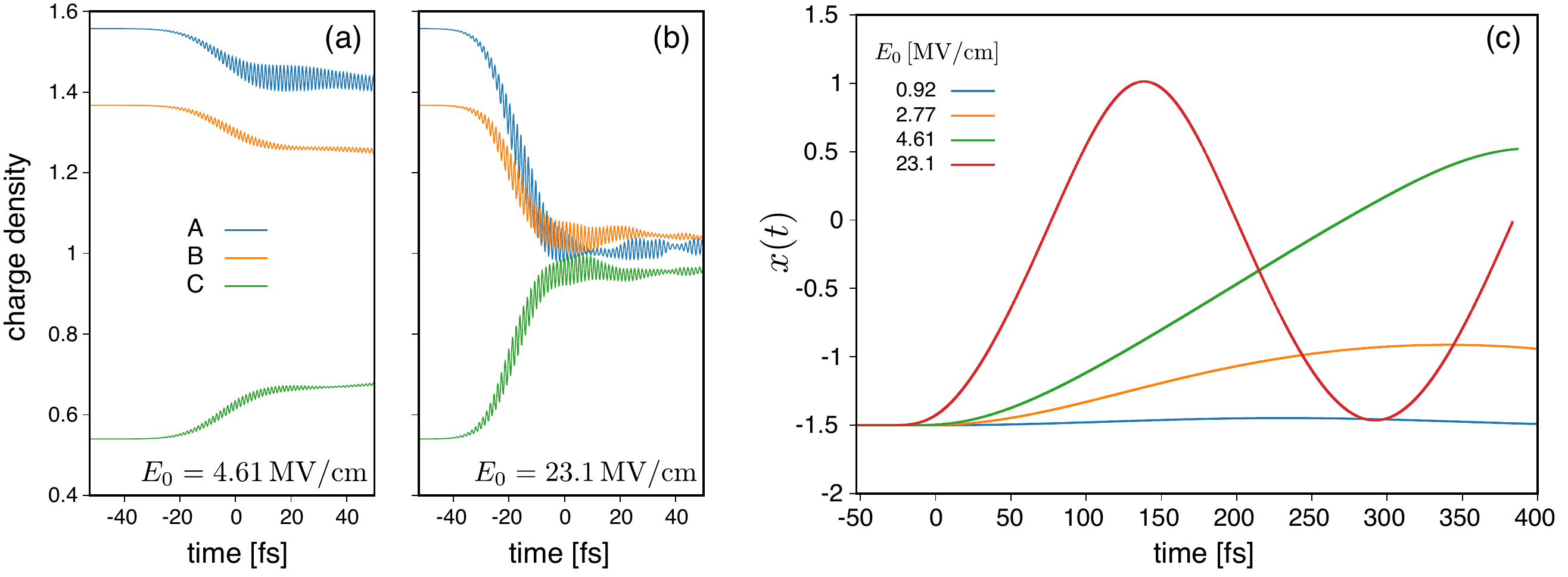}
\caption{
Time profiles of charge densities averaged over A, B, and C sites
for laser amplitudes (a) $E_0=4.61$ and (b) $23.1\mvcm$.
(c) Time profile of lattice distortion $x(t)$ for several laser amplitudes $E_0$.
}
\label{fig:stronger}
\efig


Next we discuss the energy efficiency of the photoinduced CDW melting.
We have calculated the energy absorption $\dabs\equiv E(t=0\fs)-E_\text{init}$,
and obtained $\dabs=4.6$ and $39\eV$ per two hexagrams for $E_0=4.61$ and $23.1\mvcm$, respectively.
We compare these values with the energy difference $\dcdw\equiv \euni-\eini$.
Here $\euni$ denotes the energy of a virtual uniform state
and is defined as the expectation value of
$\htot$~\eqref{eq:htotal} at $x=-1.5$ calculated for the electronic ground state of $\htot$ for $x=0$.
We obtain $\dcdw=5.8\eV$ per two hexagrams,
and it is an estimate of the energy needed to wash out the CDW order without deforming the PLD.
It is consistent with the fact that $\dabs <\dcdw$ for $E_0=4.61\mvcm$, where the CDW is suppressed but remains,
whereas $\dabs >\dcdw=0.15\dabs$ for $E_0=23.1\mvcm$, where the CDW melts down.
Thus $15\%$ of the absorbed energy is used for the CDW melting,
and the remaining $85\%$ is converted into heat.

To melt the CDW order,
the laser-electron interaction is not the only way
since the lattice motion is involved on the long time scale.
We show the lattice dynamics in Figure~\ref{fig:stronger} for some strong laser fields.
For $E_0=4.61\mvcm$, $x(t)$ hits zero, meaning that the PLD of the hexagram shape melts down
and the CDW order of electrons thereby disappears.
We note that this field amplitude is not strong enough to completely melt the CDW order on the short time scale.
However, in the long run, the laser energy absorbed by the electrons is transferred to the lattice motion,
which affects the charge density through the electron-lattice coupling.
On the long time scale, the electron-lattice cooperative dynamics becomes more important.

These electron-lattice dynamics are quite different from those
found in experimental~\cite{Iwai2007} and theoretical studies~\cite{Tanaka2010,Miyashita2010}
on charge order melting in quasi-two-dimensional organic conductors.
In these materials, the charge order is mainly produced by long-range Coulomb repulsion
and weakly assisted by an electron-lattice coupling.
Thus, the charge order can be melted efficiently by a pulse laser with little changing the lattice distortion.
See also \cite{Yonemitsu2007} for the efficiency of the charge order melting in a quasi-one-dimensional organic conductor.


\section{Discussion and Conclusion}
In this work, we have extended the tight-binding model~\eqref{eq:tb}
of the previous studies~\cite{Smith1985,Rossnagel2006}
and included the on-site Coulomb interactions~\eqref{eq:hint}.
By this extension and our mean-field analysis, we have succeeded in opening
a band gap at the Fermi level as shown in Figure~\ref{fig:bands}(e).
Thus we have established a lattice model which reproduces qualitative features
of the electronic properties of \tas.
We have further extended the model to treat the lattice distortion as a dynamical variable.
Considering a specific PLD of the hexagram shape,
we have modeled the lattice potential,
reproduced the CCDW as a thermal equilibrium state,
and formulated the coupled equations of motion for the electrons and the PLD.

We have investigated the short-time charge dynamics induced by a pulse laser on the basis of our model.
We have shown that the short-time dynamics is dominated by electrons
and the hexagram CDW order is suppressed by the laser,
while the lattice is effectively frozen.
This CDW decrease is caused at the second order of the laser electric field,
and the CDW order melts down as fast as $20\fs$ for the strong laser amplitude $E_0=23.1\mvcm$ for the $35\fs$ pulse.
We note that the necessary amplitude $E_0$ should be smaller for a longer pulse.

The lattice distortion plays an important role on the long time scale $\gtrsim\!100\fs$.
Its motion is dominated by the landscape of the free energy
instantaneously converted from the equilibrium one [see Figure~\ref{fig:lat_long}(b)].
The CDW order may be washed out on the long time scale through the coupling to the lattice motion,
even when the laser amplitude is not strong enough to melt the CDW on the short time scale.

We comment on a few open questions beyond our present study.
We have considered only one mode of the hexagram shape in lattice distortion,
but there exist many other modes giving rise to the incommensurate CDW, the nearly-commensurate CDW, and so on.
Including those relevant modes may make it possible to describe photoinduced transitions from the CCDW to the other CDW phases.
In that case, our picture of the free-energy modulation is extended in multiple dimensions and several local minima may exist.
We note that the phonon modes also work as a heat reservoir.
Our present analysis does not include a coupling to any reservoir, or dissipation,
and our results would be modified especially at long times.
We leave these important issues as future works.


\section{Materials and Methods}
\subsection{Spin-Orbit Interaction}\label{sec:soi}
We give the the expressions for $\hat{\bm{L}}_i$ and $\hat{\bm{S}}_i$:
\beq
\hat{L}^\alpha_i = \sum_{\iorb,\iorb',\isp} L^\alpha_{\iorb\iorb'}\cre_{i\iorb\isp}\ann_{i\iorb'\isp};\qquad
\hat{S}^\alpha_i = \frac{1}{2}\sum_{\iorb,\isp,\isp'} \tau^\alpha_{\isp\isp'}\cre_{i\iorb\isp}\ann_{i\iorb\isp'}
\quad (\alpha=x,y,z).
\eeq
Here 
$ \tau^\alpha$'s are the Pauli matrices
and
$L^\alpha_{\iorb\iorb'}$'s are the $3\times3$ sub-matrices for  $\iorb=$ \oz,\ \oxx,\ \oxy\
of the $L=2$ representation of the angular momentum~\cite{Sakurai1993}.
We note that only $\hat{L}^z_i \hat{S}^z_i$ is active in $\hat{\bm{L}}_i\cdot\hat{\bm{S}}_i$
within our subspace spanned by the states with $L^z_i=0,\pm2$.
The nonzero elements of $L^z$ are thus $L^z_{x^2-y^2,xy}=-L^z_{xy,x^2-y^2}=\ii$.






\vspace{6pt} 



\authorcontributions{
Conceptualization, \ike\ and \yo;
data curation, \ike;
formal analysis, \ike\ and \tsu;
funding acquisition, \ike\ and \yo;
investigation \ike;
supervision \yo;
visualization \ike;
writing -- original draft, \ike;
writing -- review \& editing, \ike, \tsu\ and \yo
}

\funding{This work was funded by JSPS KAKENHI Grants No.~JP16H06718 , JP16K05459, and JP18K13495.}

\acknowledgments{
The authors thank K. Rossnagel for providing us with the parameters for the empirical tight-binding model.
An unpublished work on a similar model by R. Fujinuma (Master's thesis, Chuo Univ., 2015) has also been used as a reference.}

\conflictsofinterest{The authors declare no conflict of interest.} 

\abbreviations{The following abbreviations are used in this manuscript:\\

\noindent 
\begin{tabular}{@{}ll}
PLD & Periodic Lattice Distortion\\
CDW & Charge Density Wave\\
CCDW & Commensurate Charge Density Wave\\
\end{tabular}}

\appendixtitles{no} 
\appendixsections{multiple} 
\appendix


\reftitle{References}




\end{document}